\documentclass[useAMS,usenatbib,usegraphicx]{mn2e}
\usepackage{times}

\def\icrit{{i_{\rm crit}}}
\title[High Inclination Planets in Multistellar Systems]{High
  Inclination Planets and Asteroids in Multistellar Systems}
\author[P. E. Verrier and N. W. Evans]
{P. E. ~Verrier$^1$\thanks{E-mail: pverrier@ast.cam.ac.uk (PEV);
    nwe@ast.cam.ac.uk (NWE)}
  and N. W. ~Evans$^1$\footnotemark[1]\\
  $^1$Institute of Astronomy, University of Cambridge, Madingley Road,
  Cambridge, CB3 0HA, United Kingdom}
\begin{document}

\date{10 July 2008}

\pagerange{\pageref{firstpage}--\pageref{lastpage}} \pubyear{0000}

\maketitle

\label{firstpage}


\begin{abstract}
  The Kozai mechanism often destabilises high inclination orbits. It
  couples changes in the eccentricity and inclination, and drives high
  inclination, circular orbits to low inclination, eccentric orbits.
  In a recent study of the dynamics of planetesimals in the quadruple
  star system HD98800 \citep{VE08}, there were significant numbers of
  stable particles in circumbinary polar orbits about the inner binary
  pair which are apparently able to evade the Kozai instability.
  
  Here, we isolate this feature and investigate the dynamics through
  numerical and analytical models. The results show that the Kozai
  mechanism of the outer star is disrupted by a nodal libration
  induced by the inner binary pair on a shorter timescale. By
  empirically modelling the period of the libration, a criteria for
  determining the high inclination stability limits in general triple
  systems is derived. The nodal libration feature is interesting and,
  although effecting inclination and node only, shows many parallels
  to the Kozai mechanism. This raises the possibility that high
  inclination planets and asteroids may be able to survive in
  multistellar systems.
\end{abstract}



\begin{figure*}
\centering
\includegraphics[width=0.9\textwidth]{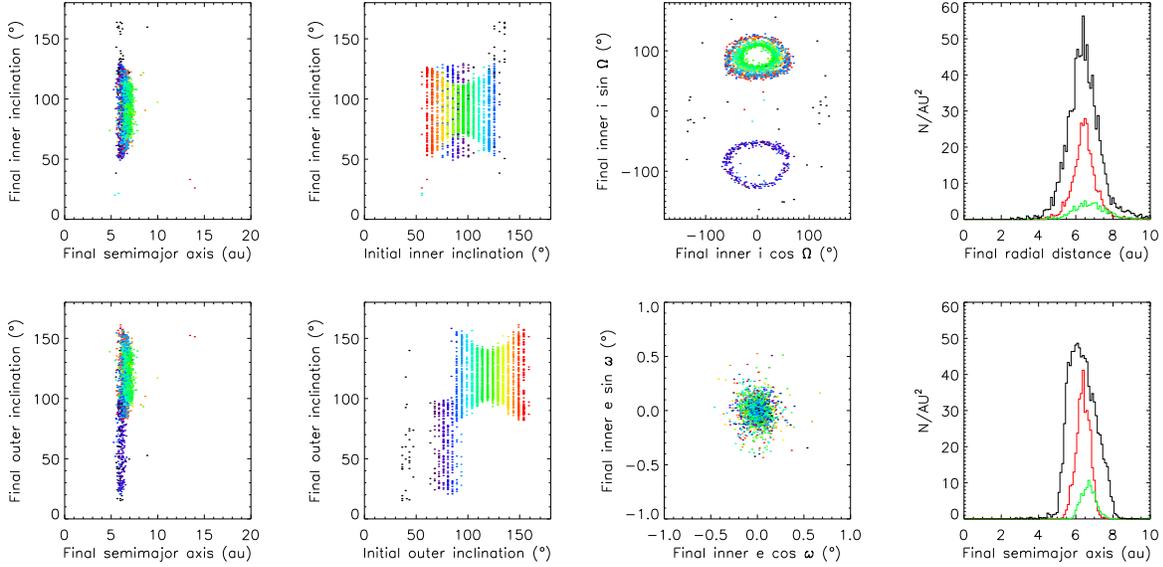}
\caption[Test particle distributions between $55^\circ$ to $135^\circ$ 
initial inclination in HD 98800]{
\label{fig:polar}
Test particle distributions for particles starting with initial
inclinations in the range $55^\circ$ to $135^\circ$ from the
simulation using orbit II in \citet{VE08}. Colour indicates initial
outer inclination, as seen in the lower centre panel. Upper and Lower
Far Left: The final semimajor axis of survivors plotted against the
final inclination with respect to the inner and outer binary. Note the
survivors are restricted to semimajor axes between 5 and 7 au. Upper
and Lower Centre Left: Initial and final inclinations of the survivors
with respect to the inner and outer binary pair.  Upper and Lower
Centre Right: The surviving distributions in the $(i \cos \Omega, i
\sin \Omega$) plane (relative to the orbital plane and pericentre of the inner binary) show two distinct populations. However, no
structure is apparent in the ($e \cos \omega, e \sin \omega$) plane.
Upper and Lower Far Right: The radial cross sections of the survivors
for all three wide orbit simulations: orbit I is in black, orbit II in
red and orbit III in green.}
\end{figure*}

\section{Introduction}
\label{sec:intro}

The Kozai instability is a well-known destabilizing mechanism of high
inclination satellites. Arnold (1990), in his book {\it Huygens \&
  Barrow, Newton \& Hooke}, reports a neat illustration due to Lidov
(1963).  Lidov discovered that if the orbit of the Moon is turned
through $90^\circ$, its eccentricity increases so rapidly under the
action of the tidal forces of the Sun that it collides with the Earth
in four years! The effect was described and analysed by Kozai (1962,
1980) in his studies of the survival of high inclination comets and
asteroids.

The Kozai mechanism is a secular effect whereby, above a critical
inclination $\icrit$ of around $40^\circ$ (depending on the system),
the inclination $i$ is coupled to eccentricity $e$ though the Kozai
constant~\citep[see e.g.,][]{Th96}
\begin{equation}
H_{\rm Koz} = \sqrt{a(1-e^2)} \cos i
\end{equation}
and is driven to vary in cycles between its initial value and the
critical value. The argument of pericentre also librates about
$\pm90^\circ$ \citep{Ko62,Th96,Ta08}.  In other words, nearly circular,
high inclination orbits are driven to high eccentricities in exchange
for lower inclination, so that in the Solar system this produces the
population of Sun-grazing comets (e.g., Stagg \& Bailey 1989).

At first glance, this suggests that planets or asteroids in high
inclination orbits in multistellar systems cannot survive for long
times.  A planet in a circumstellar orbit in a binary star system is
known to be subject to Kozai cycles on a timescale \citep{Ki98,Ta08}
\begin{equation}
\tau_{\rm{Koz}} \simeq \frac{2}{3\pi}\frac{P^2_{\rm
    bin}}{P_p}(1-e_{\rm bin}^2)^{3/2}\frac{m_A+m_B+m_p}{m_B}
\label{eq:koz}
\end{equation}
where $P_{\rm bin}$ is the period of the binary, $P_p$ that of the planet,
$e_{\rm bin}$ is the binary's eccentricity, $m_A$ and $m_B$ the stellar
masses and $m_p$ the planet's mass. Surprisingly, then, a recent study
of planetesimals in the stellar system of HD 98800~\citep{VE08} found
a long-lived, high inclination circumbinary population.

HD 98800 has four stars, in two close binaries A and B, in highly
inclined and eccentric orbits and hosts a circumbinary debris disc
around the B binary pair. The system is well approximated as a
hierarchical triple star system, as the secondary star in the A binary
is fairly small and in a close orbit. High inclination particles are
found to be in long-term stable orbits inclined by $55^\circ$ to
$135^\circ$ to the inner binary B. This is well above the critical
inclination relative to both stellar orbits, yet the orbits are not
undergoing Kozai cycles.  It is known that the Kozai effect can be
disrupted if another mechanism, such as general relativity, tidal
forces or other interactions between planets, causes orbital
precession on a shorter timescale \citep{KN91,Wu03, TR05, Ta08}.
However, the simulations in \citet{VE08} deal with the Newtonian
dynamics of test particles, so the only additional factor present is
the mutual gravitational perturbations of the stars.

This paper explains the origin of the surprising stability.  The inner
binary causes a nodal precession, instead of Kozai cycles. This
stabilises the test particles against any Kozai instability driven by
the outer star. We first present the evidence in favour of this
explanation in Section~\ref{sec:numexpt}, and then provide a more
detailed study of the nodal precession, together with the dynamics
behind it, in Section~\ref{sec:analytic}.  Finally, we summarize our
conclusions in Section~\ref{sec:conclusion}.

\begin{table}
\caption[Initial orbital parameters for the simulations of HD 98800]{
\label{tab:orbit}
The initial orbital parameters for the simulations of system HD 98800
used in \citet{VE08}. These are the orbital elements used to evolve 
the system from 1 Myr in the past to the present state, so the starting 
parameters for the inner binary are also different for each wide orbit 
case. The longitudes given are relative to the arbitrary reference direction of the simulations.}
\begin{center}
\begin{tabular}{lcccccc}
\hline
Orbital               & \multicolumn{3}{c}{Wide orbit A-B} & \multicolumn{3}{c}{Inner orbit Ba-Bb}      \\
Parameter             &  I     & II     & III                       & I & II & III                       \\
\hline
$m_1$ ($M_\odot$)     & \multicolumn{3}{c}{1.281}                  & \multicolumn{3}{c}{0.699}          \\
$m_2$ ($M_\odot$)     & \multicolumn{3}{c}{1.3}                    &  \multicolumn{3}{c}{0.582}         \\
$a$ (au)              & 61.9   & 67.6  & 78.6                      & 0.983 & 0.983 & 0.983              \\
$e$                   & 0.3    & 0.5   & 0.6                       & 0.640 & 0.790 & 0.570              \\
$i$ (${}^\circ$)      & 130.7  & 144.7 & 131.7                     & 0.0   & 0.0   & 0.0                \\
$\omega$ (${}^\circ$) & 214.1  & 323.4 &  65.9                     & 82.0  & 11.3  & 34.89              \\
$\Omega$ (${}^\circ$) & 224.5  & 309.2 &  66.6                     & 0.0   & 0.0   & 0.0                \\
$M$ (${}^\circ$)      & 166.3  & 90.8  &  16.2                     & 225.9 & 32.6  & 259.7              \\    
\hline
\end{tabular}
\end{center}
\end{table}

\section{Numerical Experiments}
\label{sec:numexpt}

In~\citet{VE08}, numerical studies of the stability of planetesimals
in the circumbinary disc of HD9800 were carried using the {\sc Moirai}
code. This is a fast symplectic code adapted for hierarchical
systems~\citep{VE07}.  We have checked that our results are not an
artefact of the numerical integration method, as similar results are
obtained with a standard \citep{NR5} Bulirsch-Stoer integration scheme
(see \citealt{Ve08}).

The stellar parameters used for the simulations are shown in
Table~\ref{tab:orbit}. We investigate three possible configurations,
labelled orbits I, II and III (see Verrier \& Evans 2008 for more
details on the rationale for the orbital parameters).  High
inclination particles are unexpectedly found to be in long-term stable
orbits inclined by $55^\circ$ to $135^\circ$ to the inner binary B,
and restricted to a narrow range of semimajor axis. These particles
are shown in Figure~\ref{fig:polar} for the case of the orbit II
parameters.  Using $\Omega$ to denote the longitude of the ascending
node, we plot the particles in the ($i \cos \Omega$, $i \sin \Omega$)
plot. This reveals distinct dynamical structures -- in particular, two
populations of stable particles. Interestingly, though, there is
little structure seen in the ($e \cos \omega$, $e \sin \omega$), where
$\omega$ is the argument of pericentre.

\begin{figure}
\begin{center}
\includegraphics[width=3.15in]{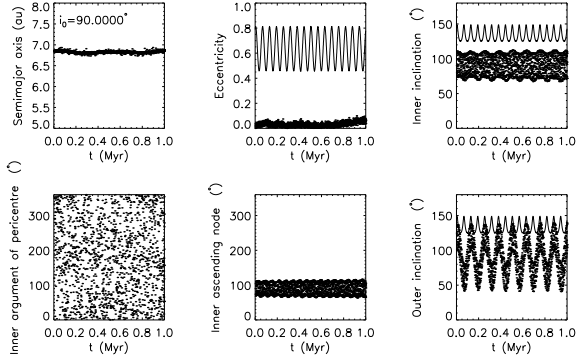}
\caption[Example orbital evolution of a particle in the stable halo]{
  The orbital evolution, relative to the inner binary orbit, of a
  particle starting near 7 au with an initial inclination of
  $90^\circ$. While the inclination variations over the Myr
  integration length are large, the eccentricity remains small and the
  particle is clearly not undergoing Kozai cycles. The inner stellar
  eccentricity and mutual inclination are overplotted as solid lines on the relevant panels.}
\label{fig:tpev}
\end{center}
\end{figure}

The orbits of all the stable particles librate about $90^\circ$ inner
(i.e. relative to the inner B binary's orbital plane)
inclination. Their nodes also librate about $\pm 90^\circ$ relative to the inner binary's pericentre, defining
the two populations seen in Figure~\ref{fig:polar}. These are also
defined by their initial node: the upper population in the ($i \cos
\Omega$, $i \sin \Omega$) plot starts at $120^\circ$ and the lower at
$240^\circ$ (relative to the simulations reference direction).  All particles starting with $\Omega = 0^\circ$ are not
stable.  Unlike the choice of initial node, there appears to be no
difference in the evolution of particles with differing pericentres or
times of pericentre passage. Interestingly, the stability of the high 
inclination particles is greater for the simulations with a lower eccentricity 
for the outer star. This is shown in the far
right panels of Figure~\ref{fig:polar}, in which there are more
surviving test particles in the case of orbit I as opposed to orbit
III. A typical orbit of one of the stable high inclination particles is
shown in Figure~\ref{fig:tpev}, and illustrates the libration of
inclination and node. There also appears to be a slight modulation of
this variation with the stellar orbital evolution, as indicated in the
Figure.

A brief investigation of the variation of stability with the stellar
parameters was achieved by integrating the stable particles for a
variety of different eccentricities and mutual inclinations of the two
stellar orbits. This revealed that the outer stellar eccentricity has
little effect on the stability, while the inner eccentricity is
crucial! As the inner orbit becomes circular, the high inclination
stability, somewhat surprisingly, decreases dramatically. The stellar
mutual inclination is only important insofar as it may cause
variations in the resulting inner eccentricity.  This would appear to
be in contradiction to the results shown in Figure~\ref{fig:polar},
where the stable region increases as the outer stars eccentricity
decreases. However, in these simulations the test particles do not
start similarly aligned with the inner star in each case (as the
original simulations start at 1 Myr in the past, where the inner star
has a different longitude for each outer eccentricity case), meaning
that in the Orbit I case the particles are closer to the libration
island, while in the Orbit III case they are further away and hence
less stable. This effect is discussed in more detail in
Section~\ref{sec:ertbp}.

\begin{figure}
\begin{center}
\includegraphics[width=3.15in]{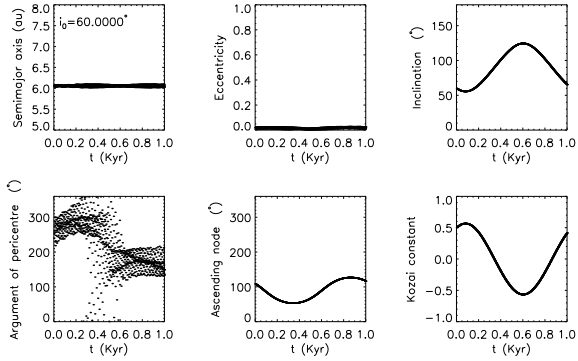}
\caption[Example circumbinary test particle evolution]{
  The typical short term evolution of a particle when the outer star
  is removed. Orbital elements are relative to the inner binary's
  orbit, using the star's periastron as the reference direction. }
\label{fig:noaev}
\end{center}
\end{figure}

The orbital evolution of particles due to each stellar orbit is also
of interest. Replacing the inner binary as a single star results in
the stable particles now undergoing Kozai cycles and becoming rapidly
unstable at very high inclinations. However, removing the outer star
instead does not have similar results. In this case, the particles
remain stable and still librate about $90^\circ$ inclination and
longitude of the ascending node, as shown in
Figure~\ref{fig:noaev}. The librations have a period of the order of
Kyrs which increases with particle semimajor axis and an amplitude
that decreases with semimajor axis.

This nodal libration could be suppressing the Kozai cycles caused by
the outer star, and could explain the distribution of the stable
particles, if the period is shorter than that of the cycles. At 3 au,
$\tau_{\rm{Koz}} \sim 3$ Kyr and at 8 au $\tau_{\rm{Koz}} \sim 1.5$
Kyr, while the nodal libration period at 3 au is about 1 Kyr, indeed
shorter than $\tau_{\rm{Koz}}$, and at 8au is about 2 Kyr, longer than
$\tau_{\rm{Koz}}$, supporting the idea.

In summary, the numerical evidence suggests that the inner binary star
causes the nodes and inclinations of particles within about 7 au to
librate on a shorter timescale than the Kozai cycles induced by the
outer star, thereby suppressing the mechanism and stabilising their
orbits.  The simulations also indicate the period of the librations
increase as the inner binary's eccentricity decreases, explaining the lack of
stability of the particles in the triple system with a circular inner
binary.  This nodal libration is itself a very interesting phenomenon,
as is the lack of Kozai cycles.

\section{The Nodal Libration Mechanism}
\label{sec:analytic}

\subsection{The Elliptic Restricted Three-Body Problem}
\label{sec:ertbp}

The stellar system is now reduced to that of the inner binary
only. The outer binary can be neglected, as we have just demonstrated
that the nodal libration takes place in its absence.  This arrangement
is now identical to the elliptic restricted three-body problem.  The
circular restricted three-body problem is the limit $e_{\rm bin
}=0$. It is most commonly met under the guise of the Copenhagen
problem, in which the two bodies have equal mass.  There has been
substantial effort on the classification of the orbits in the
Copenhagen problem~\citep[see e.g.,][]{Co67}, but the eccentric case
has been much less well studied.

Simulations are run for 10 Kyrs, as the short term behaviour is of
interest.  Since the initial node of the test particles appears to be
an important factor in the particle's orbital evolution, a set of
simulations were run looking at a much wider range of this parameter.
The underlying stellar parameters are those corresponding to orbit II,
whilst the initial test particle grid is shown in
Table~\ref{tab:tpgrid}.

\begin{figure*}
\begin{center}
\includegraphics[width=0.9\textwidth]{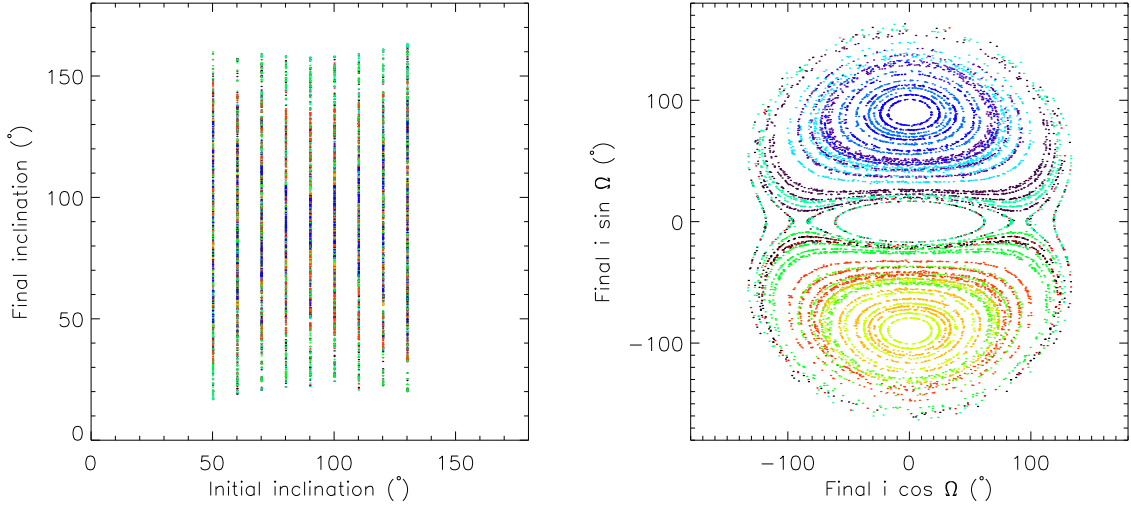}
\caption[Inclination variations and surface of section for particles
around the inner binary star of HD 98800]{ The inclination variations
  and surface of section for test particles around the inner binary
  only. The orbital elements are all relative to the orbit of this
  binary (i.e. the node is from the pericentre of the binary's
  orbit). Colour indicates initial longitude of ascending node, with
  black through blue through green through yellow through red
  representing $0^\circ$ to $360^\circ$. The librationcd islands about
  $\Omega=\pm90^\circ$ and $i=90^\circ$ are clear on the left-hand
  plot of $i \cos \Omega$ versus $i \sin \Omega$.}
\label{fig:hdtest}
\end{center}
\end{figure*}

\begin{figure*}
\begin{center}
\includegraphics[width=0.9\textwidth]{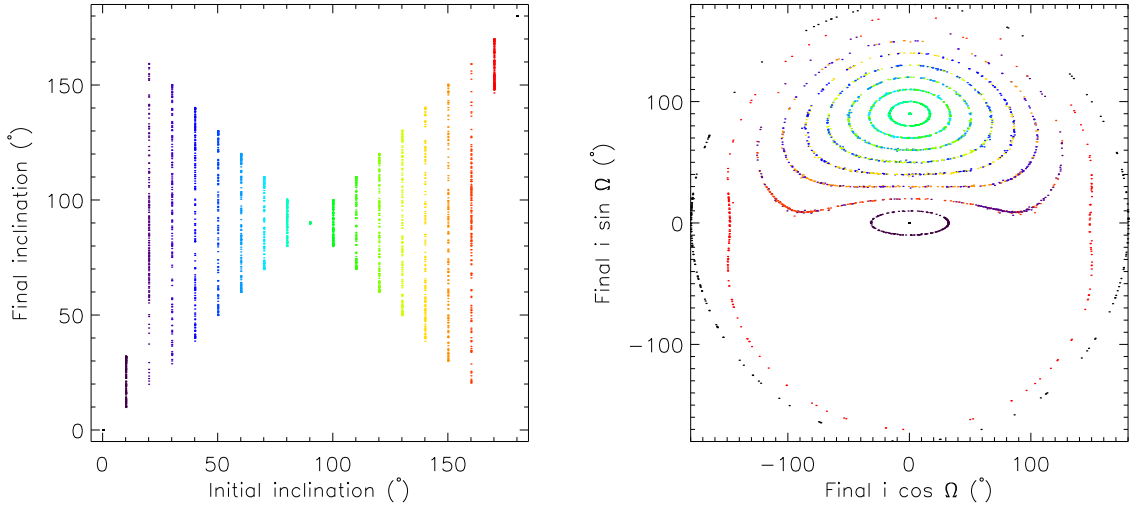}
\caption[Inclination variations and surface of section for particles
around the inner binary starting close to one of the libration
centre]{ The inclination variations and surface of section for
  particles around the inner binary of HD 98800, but now spaced evenly
  about $\Omega = 90^\circ$ and for inclinations in the range
  $0^\circ$ to $180^\circ$. Colour now indicates initial inclination,
  as shown in the left-hand plot. The initially low inclinations
  remain in the range $0^\circ$ to $30^\circ$ and are seen in the
  small libration island about the centre of the right-hand plot.}
\label{fig:hdalti}
\end{center}
\end{figure*}

The resulting distributions in inclination and the ($i \cos
\Omega$, $i\sin \Omega$) plane are shown in
Figure~\ref{fig:hdtest}. Once again the two libration islands are
clear and are at about $i=90^\circ$ and $\Omega=\pm90^\circ$ relative
to the pericentre line of the binary star. Figure~\ref{fig:hdalti}
shows these islands in more detail for a similar simulation that now
has particles spaced exactly about these libration centre (i.e. the
binary's pericentre has been set to lie along the simulation reference
direction) and from $0^\circ$ to $180^\circ$ inclination. Particles in
the centre of the islands remain there, and the initially low
inclination particles also remain in low inclination orbits, librating
about the centre of the ($i \cos \Omega$, $i\sin \Omega$) plane while
the retrograde particles circulate.  This libration and critical angle
is in striking similarity to Kozai mechanism, where similar libration
islands are seen, but for $\omega$. Here, however, the argument of
pericentre still rapidly circulates, and no structure is apparent in
the ($e \cos \omega$, $e\sin\omega$) plane.

A particle's proximity to the centre of the libration island is
controlled by its initial node and inclination: higher inclinations
and nodes place the particle further from the centre and increase the
period of the variation (which makes them less stable against the
Kozai cycles of the outer star). In addition, particles that do
circulate in both these simulations show larger eccentricity
variations and in some cases are not stable.

\begin{table}
\caption[The test particle grid used in the simulations in Section~\ref{sec:analytic}]{
The test particle grid used in the simulations shown in 
Figures~\ref{fig:hdtest} and \ref{fig:hdalti}.}
\label{tab:tpgrid}
\begin{center}
\begin{tabular}{lccc}
\hline
Orbital Elements                 &    Min  &   Max   &  Step Size \\
\hline
$a_{\rm{tp}}$ (au)                &   3.0   &  10.0   &  0.2       \\
$e_{\rm{tp}}$                     &   \multicolumn{3}{c}{0}        \\
$i_{\rm{tp}}$ (${}^\circ$)        &   50    &   130   &  10        \\
$\omega_{\rm{tp}}$ (${}^\circ$)   &   \multicolumn{3}{c}{0}        \\
$\Omega_{\rm{tp}}$ (${}^\circ$)   &   0     &   300   & 30         \\ 
$M_{\rm{tp}}$ (${}^\circ$)        &   0     &   240   &  120       \\
\hline
\end{tabular}
\end{center}
\end{table}

This explains why in the original HD 98800 simulations only those
particles with nodes (relative to the simulations arbitrary reference direction) near $120^\circ$ or, to a lesser extent,
$240^\circ$ survived: the other nodal value of $0^\circ$ placed
particles in the circulation regime, resulting in destabilising
increases in eccentricity.  For the orbit I case, far more particles
are stable as the star's longitude means some initially lie very close
to the libration centres, while orbit III particles are less stable as
they are further away from it.
\begin{figure}
\begin{center}
\includegraphics[width=2in]{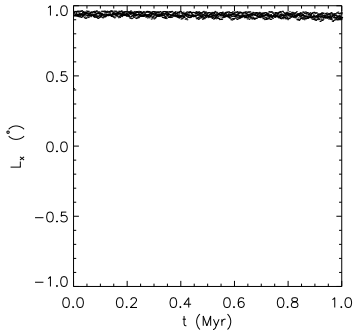}
\caption{The conservation of the angular um component, $L_{\omega_*}$,
  along the inner binary's line of apses for the same particle as
  Figure~\ref{fig:tpev}.}
\label{fig:angmo}
\end{center}
\end{figure}

\subsection{An Approximate Integral of Motion}

The surfaces of section (Figures~\ref{fig:hdtest}
and~\ref{fig:hdalti}) suggest that there is an approximate integral of
the motion for the circumbinary test particles. From the simulation
data, we can verify that this is the case. The constant is the test
particle's component of angular momentum along the direction of the
line of apses of the binary star's orbit, which is
\begin{equation}
L_{\omega_*} = h \sin i_{\rm tp} \sin \Omega_{\omega_*}
\end{equation}
where $i_{\rm tp}$ is the particle's inclination relative to the plane of
the binary's orbit, and $\Omega_{\omega_*}$ is the particle's node in
the same frame of reference and relative to the pericentre of the
binary's orbit. An example of the conservation of this quantity is
given in Figure~\ref{fig:angmo} for the test particle from
Figure~\ref{fig:tpev}. The libration seen in the inclinations and
nodes of the test particles corresponds to a simple constant
precession of their orbital planes about this axis.

This suggests that the physical reason for the test particle's orbital
evolution is a torque on the orbit due to the component of the
gravitational force from the binary parallel to the axis of the line
of apses, labelled as the direction $x_{\omega_*}$. This offers an
explanation for the increased stability with eccentricity. As the
binary's orbit becomes more eccentric it approaches a straight line
along the direction $x_{\omega_*}$, resulting in a greater torque and
hence faster precession. As the binary's eccentricity decreases, the
torque due to forces along the star's minor axis (i.e. parallel to the
$y_{\omega_*}$ direction, using the previous notation) will increase,
which eventually acts to suppress the precession about the
$x_{\omega_*}$ axis.

\begin{figure*}
\begin{center}
\includegraphics[width=6in]{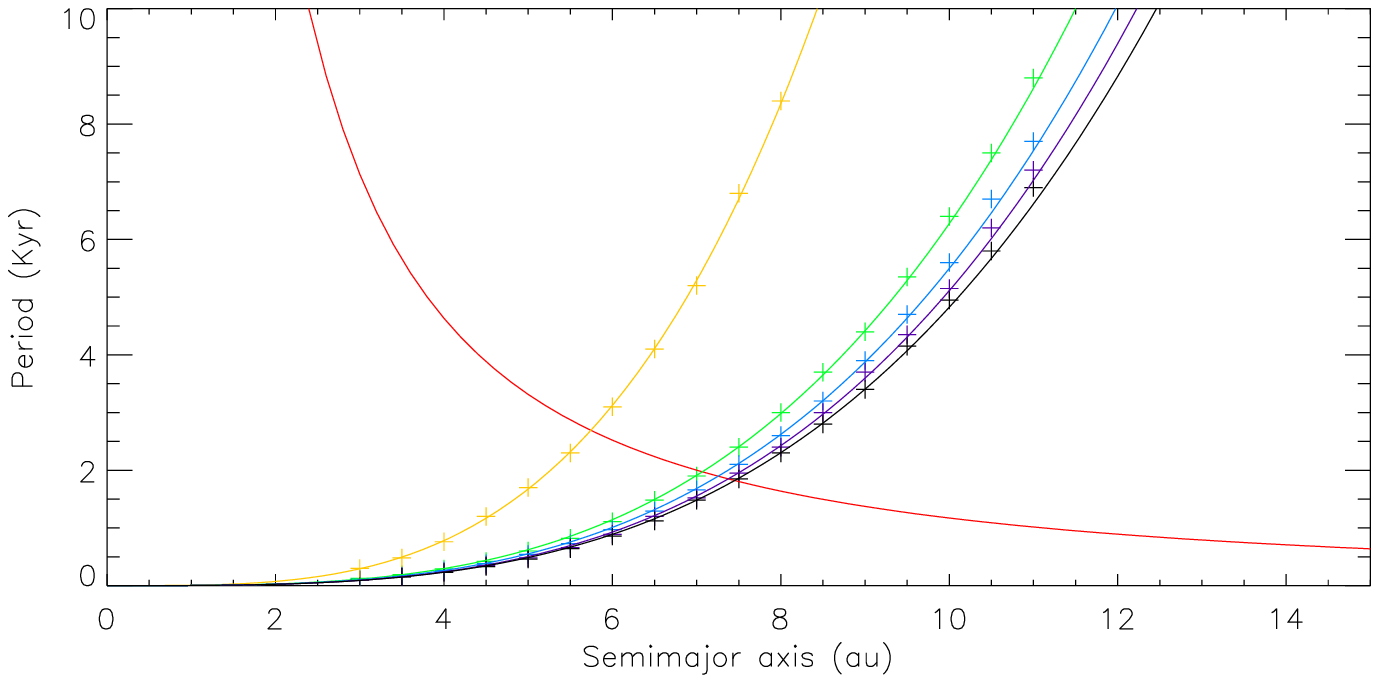}
\caption[The period of the inclination variation as a function of
semimajor axis]{ The period of the inclination variations plotted as a
  function of semimajor axis, for test particles around the inner
  binary of HD 98800. The prograde inclinations are shown only, as the
  retrograde cases are almost identical. $50^\circ$ initial
  inclination is in green, $60^\circ$ in blue, $70^\circ$ in purple
  and $80^\circ$ in black and the fits to the data are shown as solid
  lines. The Kozai timescale of the orbit II outer binary star, shown
  in red, intersects the period variation lines near the outer edge of
  the orbit II stable halo. The period of the inclination variations
  is also shown in orange for an inner binary with eccentricity
  $e=0.45$, the minimum value obtained by the stars in the HD 98800
  orbit II triple system, and can be seen to be much longer than the
  higher eccentricity case.}
\label{fig:fits}
\end{center}
\end{figure*}

\subsection{Stability Boundaries}

The period and amplitude of the librations vary with initial
inclination, node and semimajor axis and the shortest period and
minimum amplitude occur for particles that start close to the
libration centre. The point at which the period of these variations
equals the Kozai timescale of the outer star defines the outer
stability boundary. This can be modelled empirically by fitting the
period of the inclination variations.

The nodal librations are best defined for initial nodes of
$\Omega=90^\circ$.  To fit the period, sets of simulations are run
with a similar grid of test particles as shown in
Table~\ref{tab:tpgrid}, but with $\Omega$ and $M$ fixed at $90^\circ$
and $0^\circ$ respectively and the semimajor axis varied from 3 to 15
au now in steps of 0.5 au.  The period of the inclination variations
are shown in Figure~\ref{fig:fits} as a function of initial semimajor
axis for the different inclinations. A power law fits each inclination
case very well. This gives the following empirical law
\begin{equation}
P = N a_{\rm tp}^n \,\rm{Kyr}
\label{eq:per}
\end{equation}
where $a_{tp}$ is the test particle semimajor axis and the associated
best fit parameters and uncertainties are given in Table~\ref{tab:wyn}

The Kozai timescale for HD 98800 is also shown in
Figure~\ref{fig:fits}, and meets the $80^\circ$ inclination line at
7.4 au. This matches up very well with the outer edge of the high
inclination stable test particles for HD 9880 (see the left panels of
Figure~\ref{fig:polar}). Therefore, this accurately describes the
stability boundary for HD 98800 and can be used as a method of finding
a general high inclination stability limit. It should also be noted
that the minimum eccentricity of $0.45$ obtained by the inner binary
in the three star case results in longer period variations, over
plotted on Figure~\ref{fig:fits}, which probably causes the slightly
diffuse outer edge of the halo seen in Figure~\ref{fig:polar}.

\begin{table}
\caption{Fitted parameters for the period law given in eqn~(\ref{eq:per})}
\label{tab:wyn}
\begin{center}
\begin{tabular}{lccc}
\hline
Inclination & $N$ & $n$& $\chi^2$ \\\hline
$50^\circ$ & $0.0029\pm 0.0008$ & $3.33\pm 0.14$ & 0.32\\ 
$60^\circ$ & $0.0027\pm 0.0007$ & $3.32\pm 0.14$ & 0.26\\
$70^\circ$ & $0.0023\pm 0.0006$ & $3.33\pm 0.14$ & 0.34 \\
$80^\circ$ & $0.0024\pm 0.0006$ & $3.31\pm 0.14$ & 0.45 \\
$130^\circ$ & $0.0026 \pm 0.0007$ & $3.40\pm 0.14$ & 0.26\\
$120^\circ$ & $0.0024 \pm 0.0006$ & $3.36\pm 0.14$ & 0.15\\
$110^\circ$ & $0.0022 \pm 0.0006$ & $3.36\pm 0.14$ & 0.23\\
$100^\circ$ & $0.0022 \pm 0.0006$ & $3.35\pm 0.14$ & 0.08
\end{tabular}
\end{center}
\end{table}
\begin{figure*}
\begin{center}
\includegraphics[width=\textwidth]{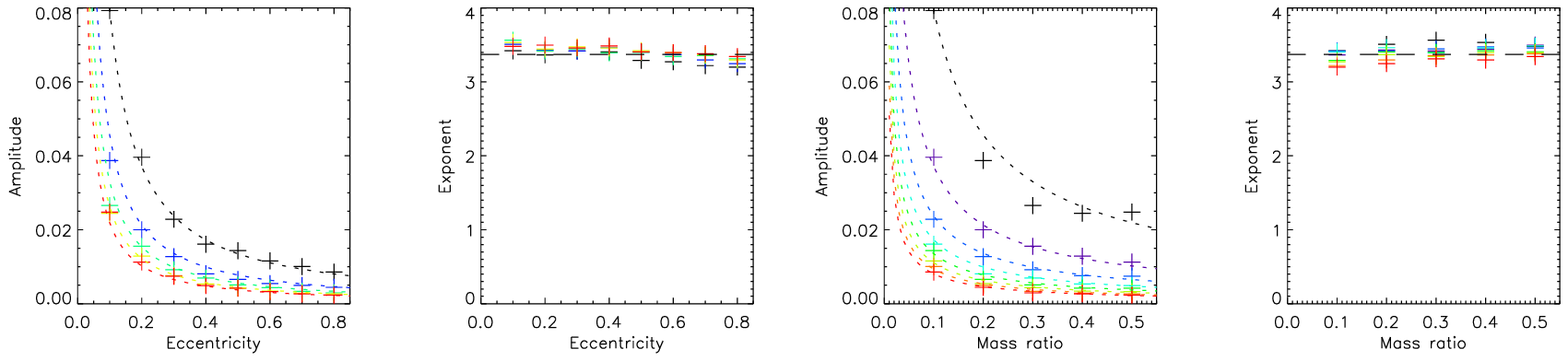}
\caption[Variation of the fitted parameters with binary mass ratio and eccentricity]{
\label{fig:hdemu}
The variation of the amplitude and exponent of the fitted power law
with mass ratio and eccentricity for the general binary case. In the
top two panels different mass ratios are indicated by the colours
purple (0.1), blue (0.2), green (0.3), orange (0.4) and red (0.5). In
the bottom two panels, different eccentricities are indicated by the
colours black (0.1), purple (0.2), blue (0.3), light blue (0.4), green
(0.5), yellow (0.6), orange (0.7) and red (0.8). The fitted lines are
also shown.}
\end{center}
\end{figure*}

We now carry out the same calculation for sets of test particles in
general binary systems.  This empirical model can then be used in
conjunction with the outer star's Kozai timescale to place limits on
the high inclination stability in a general hierarchical triple system.

Simulations show that the libration islands decrease in size as the
eccentricity decreases, whilst the critical inclination for the
librations to occur increases. The mass ratio of the binary, $\mu =
m_{Bb}/(m_{Ba}+m_{Bb})$, also has an effect, but not on the geometry
of the libration islands. Instead, as this ratio decreases from 0.5,
the particles start to escape from the simple libration patterns
previously seen. To derive our empirical fit, sets of simulations are
again run with similar test particle grids, but now with only one
inclination of $85^\circ$ to find the longest possible period of
variations. The stellar mass ratio is varied from 0.1 to 0.5 in steps
of 0.1 and the eccentricity from 0.1 to 0.8 in steps of 0.1. The
binary was generalised to a separation of 1 au, total mass 1
$\rm{M}_\odot$ and with all longitudes equal to zero. The parameters
to the power law were found to fit
\begin{equation}
  \frac{P (\rm{Kyr})}{P_{\rm bin} (\rm{yr})} \simeq 0.001
  e^{-1.1}\mu^{-0.8}\left(\frac{a}{a_{\rm bin}}\right)^{(3.37\pm0.06)}.
\label{eq:fit}
\end{equation}
This is expected to scale with semimajor axis, as seen in the
coplanar limits in \citet{VE07}. To derive this empirical formula, the
powers of $e$ and $\mu$ have been fixed at the values shown in the
fitting. The amplitude has been fitted (excluding the $e=0.1$ case,
which seems poorly modelled by this formula, most likely as it is
approaching the constant inclination regime), but has an associated
error of 0.01.  Despite this the equation provides a good fit to the
data and a low $\chi^2$ value, as shown in Figure~\ref{fig:hdemu}.

\section{Conclusions}
\label{sec:conclusion}

This paper presents a foray into an area of dynamics that has not been
extensively studied, namely small particles around a highly eccentric
binary star in a hierarchical triple system. At outset, such particles
might be expected to be disrupted by the Kozai instability.

Here, we have demonstrated the existence of stable, high inclination
circumbinary test particles. They owe their stability to the high
eccentricity of the inner binary. This is somewhat surprising, as it
goes against every expectation of planetary dynamics. The inner
binary, instead of inducing Kozai cycles, causes smooth inclination
variations and nodal precession for certain initial longitudes. This
suppresses Kozai cycles that would otherwise occur due to the
outer star in the hierarchical triple.

An analytical theory to explain these variations is desirable.
However, the stars are very eccentric, their mass ratio is high, and
the test particle's inclination is also large, leaving no obvious
small parameters for the standard perturbation expansions of celestial
mechanics. The semimajor axis ratios are not very small, but do not
explain why existing secular theories (e.g.  \citealt{Ko62,FKR00}) are
unable to describe the dynamics. The libration islands, inclination
variations and critical initial inclination are surprisingly familiar
to, but contrasting with, the dynamics of the Kozai mechanism. The
Kozai mechanism occurs if $\dot\omega \approx 0$, whereas the nodal
libration appears to be a similar process occurring when $\dot\Omega
\approx0$.  There exists an approximate integral of motion -- namely,
the test particle's component of angular momentum along the direction
of the line of apses of the binary star's orbit. This also is
analogous to the Kozai mechanism, which has an approximate integral of
motion.

The high inclination libration feature may have important consequences
for planetary stability in circumbinary orbits. Many stellar members
of triples have exchanged into the system, so high mutual planet-star
inclinations are very likely. If there are regions of stability then
the outlook for planetary systems in these environments is more
promising than previously thought.

\section*{Acknowledgments}

PEV acknowledges financial support from the Science and Technology
Facilities Council. We thank John Chambers for helpful comments on the
paper.



\label{lastpage}

\end{document}